\documentclass[9pt,twocolumn,twoside]{opticajnl}
\journal{opticajournal} % use for journal or Optica Open submissions

\setboolean{shortarticle}{true}
\usepackage{lineno}
\title{Fuzzy Clustering for Low-Complexity Time Domain Chromatic Dispersion Compensation Scheme in Coherent Optical Fiber Communication Systems}

\author[1]{Wenkai Wan}
\author[1,*]{Aiying Yang}
\author[1]{Peng Guo}
\author[1]{Zhe Zhao}
\author[1]{Tianjia Xu}
\author[1]{Jinxuan Wu}
\author[1]{Zhiheng Liu}

\affil[1]{Key Laboratory of Photonics Information Technology, Ministry of Industry and Information Technology, School of Optics and Photonics, Beijing Institute of Technology, Beijing 100081, China}
\affil[*]{yangaiying@bit.edu.cn}

\begin{abstract}
Chromatic dispersion compensation (CDC), implemented in either the time-domain or frequency-domain, is crucial for enhancing power efficiency in the digital signal processing of modern optical fiber communication systems. Developing low-complexity CDC schemes is essential for hardware implemention, particularly for high-speed and long-haul optical fiber communication systems. In this work, we propose a novel two-stage fuzzy clustered time-domain chromatic dispersion compensation scheme. Unlike hard decisions of CDC filter coefficients after determining the cluster centroids, our approach applies a soft fuzzy decision, allowing the coefficients to belong to multiple clusters. Experiments on a single-channel, single-polarization 20Gbaud 16-QAM 1800 km standard single-mode fiber communication system demonstrate that our approach has a complexity reduction of 53.8\% and 40\% compared with clustered TD-CDC and FD-CDC at a target Q-factor of 20\% HD-FEC, respectively. Furthermore, the proposed method achieves the same optimal Q-factor as FD-CDC with a 27\% complexity reduction.
\end{abstract}

\setboolean{displaycopyright}{false} % Do not include copyright or licensing information in submission.

\begin{document}

\maketitle

\paragraph*{\large Introduction}
In digital coherent receivers, chromatic dispersion compensation (CDC) is performed through digital signal processing (DSP). CDC can be implemented in both the time-domain (TD-CDC) and frequency-domain (FD-CDC). Fast Fourier transform (FFT)-based CD equalization is commomly adopted in the existing systems \cite{spinnler2010equalizer,poggiolini2009evaluation}. Due to the high-complexity, it consums around 20\% power in the receiver and becomes the major challange of coherent optical communication because of the high power consumption\cite{minkenberg2021co,xing2024low}.
Also, its limited interaction with other time-domain modules restricts the potential reduction of energy consumption \cite{zhu2012optimal}. Moreover, the time-frequency conversion process in FFT/IFFT introduces higher latency compared to TD-CDC, which is particularly undesirable in data center communications and 5G networks \cite{ohlen2016data}. Driven by these challanges, the development of reduced-complexity TD-CDC methods has become a key issue of research.

One effective approach is the clustering of time-domain filter taps, which can be applied not only in intensity-modulated direct detection (IM/DD) systems \cite{huang2023low}, but also in coherent detection systems. Recently, a Time-Domin Clustered Equalizer (TDCE) was proposed, leveraging the tap redundancy phenomenon to reduce the complexity of the time-domain equalizers \cite{gomes2024geometric}. When combined with machine learning techniques, further complexity reduction is achievable. 
This approach is effective when the fiber transimission length is less than 480 km, as shown in Fig.4(f) of Ref.\cite{gomes2024geometric}. For longer transimission fiber length, the tap overlapping effect becomes less pronounced, making the complexity reduciton less significant compared to state-of-the-art frequency-domain equalizers.

In this letter, we propose a novel two-stage fuzzy clustered TD-CDC algorithm. In this approach, cluster centroids of time-domain filter taps are determined using the K-means clustering method, followed by a soft decision process for classifying each filter coefficient. The soft decision is based on the Euclidean distances between each coefficient point and its two nearest centroids in the complex plane. If the shorter distance is below a certain threshold, indicating strong adherence to the nearest centroid, the coefficient is classified into one cluster. Otherwise, it is classified into two nearest clusters. We experimentally demonstrate a complexity reduction of 40\% and 53.8\% compared with FD-CDC and clustered TD-CDC, respectively, in a single carrier 20Gbaud 1800 km coherent optical fiber transimission system. Additionally, the proposed method achieves the optimal performance as the state-of-the-art FD-CDC while reducing complexity by 27\%.

\paragraph*{\large Principle of the fuzzy clustered time-domain CDC.} Chromatic dispersion equalization can be implemented in time domain using a linear complex-valued FIR filter \cite{savory2008digital}, derived from the inverse Fourier transform of the the linear transfer function. 

Each filter tap $g(k)$ represents a point on a circle in the complex plane \cite{gomes2024geometric}, which can be given by \cite{savory2008digital}:
\begin{equation}
  g(k) = \sqrt{\frac{j c T^2}{D \lambda^2 z}}\exp(-j\frac{\pi c T^2}{D \lambda^2 z}k^2)
  \label{eq:FIR coefficients}
\end{equation}
where $z$ is the fiber length, $k$ is the filter tap index, $j$ is the imaginary unit, $D$ is the fiber chromatic dispersion coefficient, $c$ is the speed of light, $\lambda$ is the central wavelength of the transmitted optical wave, and $T$ is the sampling period. To take into consideration of the Nyquist sampling frequency to avoid aliasing, the maximum number of filter taps $N_{max}$ is expressed as:
\begin{equation}
  N_{max} = 2 \times \left \lfloor \frac{\left | D \right |\lambda^2z}{2cT^2} \right \rfloor + 1
\end{equation}
Since the impulse response of the FIR filter is symmetric about its center, employing this distributive property enables a reduction of around 50\% in complex multiplcation operations \cite{spinnler2010equalizer}. However, for large accumulated chromatic dispersion, such as in uncompensated long-haul fiber links, it is still far from being implement-efficient. 

\begin{figure}
  \centering
  \includegraphics[width=0.75\linewidth]{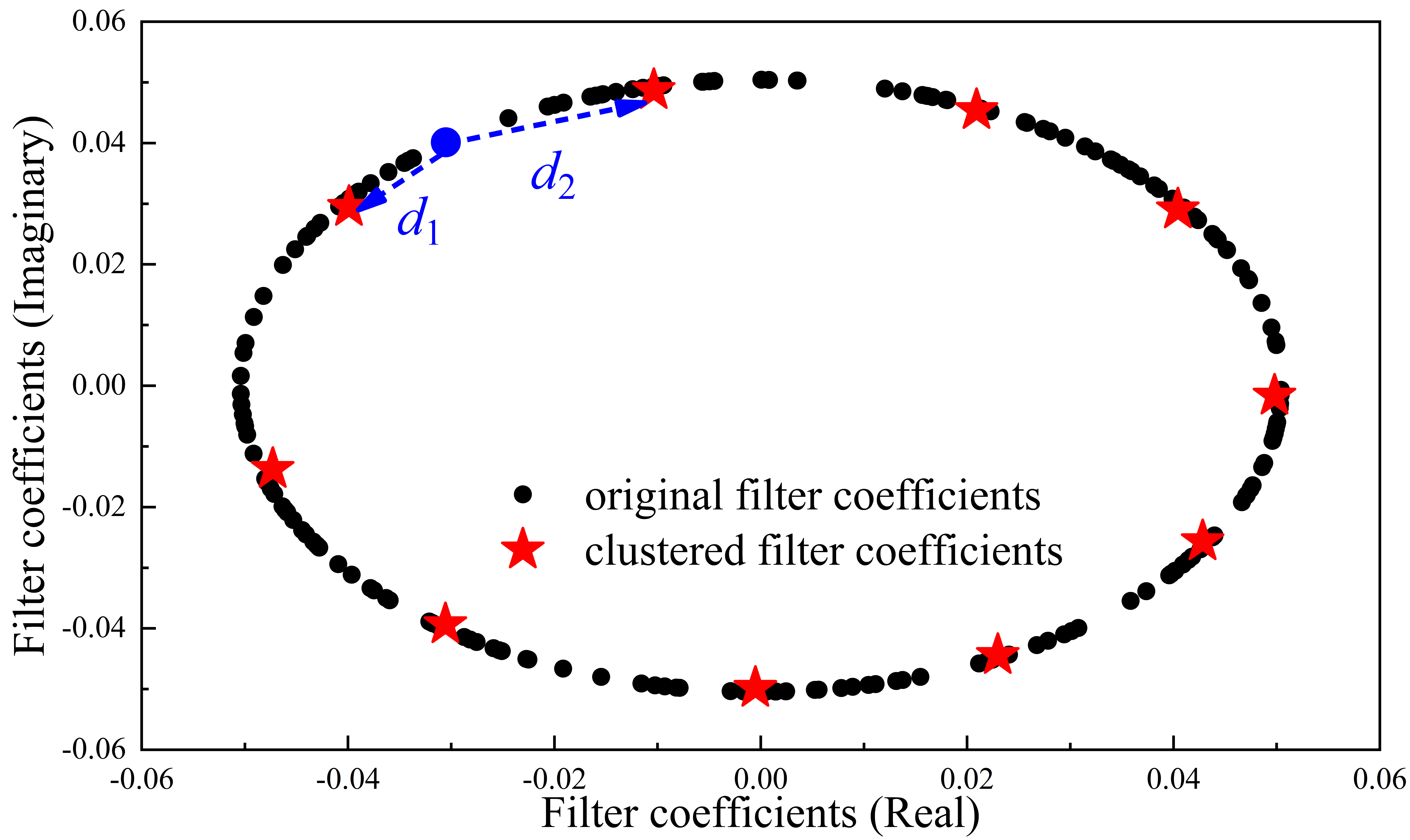}
  \caption{The distribution of filter taps in the complex plane for the time-domain CDC versus the clustered one. Blue dots indicate the filter taps involved in fuzzy clustering. $d_1$ is the distance to the nearest centroid, and $d_2$ is the distance to the second-nearest centroid.}
  \label{fig:filter coefficients distribution.}
\end{figure}

Upon closer inspection of Eq.(\ref{eq:FIR coefficients}), it can be observed that it represents a rotating vector with constant amplitude and a phase that varies with the tap index $k$. As the absolute phase increases for different values of $k$, many phase values will be repeated or become nearly identical on the complex plane, as phase values exceeding 2$\pi$ correspond to multiple rotations around the complex circle \cite{gomes2024geometric}. 
As shown in Fig.\ref{fig:filter coefficients distribution.}, this tap redundancy enables the grouping of multiple filter taps into clusters, where each cluster is represented by a single tap. Therefore, a scheme referred to as clustered time-domain CDC was proposed in \cite{gomes2024geometric}. By first calculating the summation of the samples accociated with the grouped filter taps within each cluster, and then multiplying the resultant sum by the corresponding clustered filter taps, the number of complex multiplcation operations can be reduced. The simplified chromatic dispersion equalization is experessed as:
\begin{equation}
  y(n) = \sum_{k=0}^{N_\text{c} - 1}x_\text{s}(k)g_\text{c}(k)
\end{equation}
where $n$ is the output sample index, $g_\text{c}(k)$ represents the clustered filter taps, $x_\text{s}$ represents the summation of input samples associated with the same cluster of filter taps for each $n$, and $N_\text{c}$ denotes the total number of complex value clusters. 
This scheme offers a good performance-complexity tradeoff with the transimission fiber lenght up to 480 km, where the taps overlapping effect is more pronounced. However, at long-haul distances with large dispersion accumulation, the phase values distribution in the complex plane become more spread out \cite{gomes2024geometric}, and using fewer clusters leads to a greater performance penalties.

\begin{figure}
  \centering
  \includegraphics[width=\linewidth]{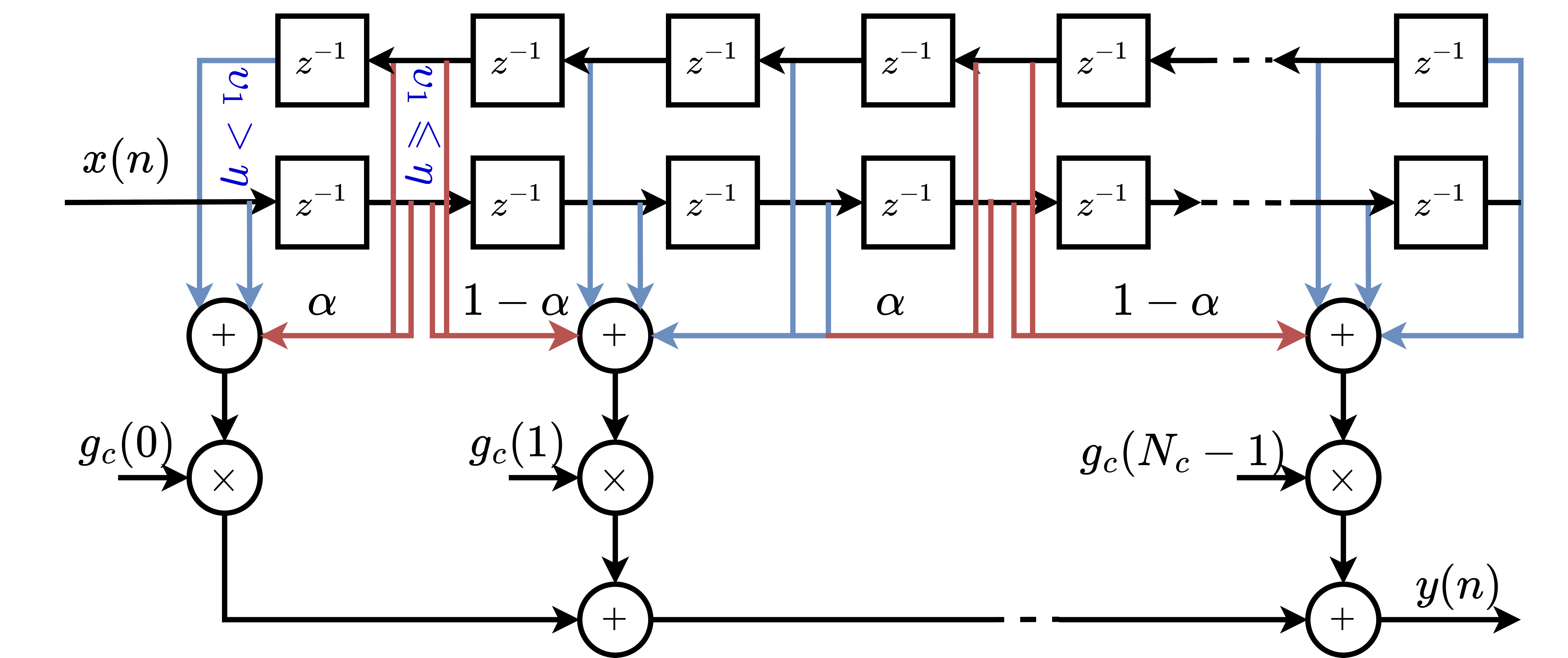}
  \caption{Proposed fuzzy clustered time-domain chromatic dispersion compensation filter architecture.}
  \label{fig:fuzzy cluster FIR architecture.}
\end{figure}

In the hard clustering algorithm, such as K-means clustering algorithm, each filter tap is strictly assigned to a single cluster, with membership values equaling either 0 or 1. As a result, the filter taps  farther from its center of a cluster contributes dominantly to clustering-induced errors. To mitigate this issue, we propose a soft clustering approach based  on fuzzy set theory \cite{zimmermann2010fuzzy}, where a filter tap can belong to different clusters with different membership values $\upsilon_i$, with $\sum_{i}\upsilon_i = 1$. This method is implemented by a two-stage process. First, similar to clustered TD-CDC, we employ the K-means clustering method to group the original tap coefficients of the FIR filter and determine the cluster centroids. Next, we calculate the distances between each tap coefficient and its two nearest cluster centroids, denoted as $d_1$ and $d_2$ (where $d1 \leqslant d2$), as shown in Fig.\ref{fig:filter coefficients distribution.}. After normalizing the inverse of $d_1$ and $d_2$, we obtain $\upsilon_1 = d_2 / (d_1 + d_2 )$ and $\upsilon_2  = d_1 /(d_1 + d_2)$ (with $\upsilon_1 \geqslant \upsilon_2$). We then introduce a threshold factor $\eta$. Coefficients with $\upsilon_1 > \eta$ are reasonably considered to have stronger adhesion to a particular cluster centroid and are rigidly classified. For coefficients with $\upsilon_1 \leqslant \eta$, they should be associated with the two closest clusters. To avoid introducing additional multiplcations, their membership values $\upsilon_1$ and $\upsilon_2$ are optimized to a fixed pair of weights,  $\alpha$ and $1 - \alpha$. This approach, which we call fuzzy clustered time-domain CDC, can be expressed as:
\begin{equation}
  y(n) = \left[\sum_{k = 0}^{N_\text{c} - 1}(x_{\text{s}}^{\text{NF}}(k) + \alpha x_{\text{s, 1}}^{\text{F}}(k) + (1 - \alpha)x_{\text{s, 2}}^\text{F}(k))\right]g_\text{c}(k)
\end{equation}
where $x_{\text{s}}^{\text{NF}}$ represents the sum of input samples linked to the same cluster of filter taps that are not subjected to fuzzy clustering, while $x_{\text{s}}^{\text{F}}$ represents the sum of input samples associated with the same cluster of filter taps that undergo fuzzy clustering. The subscripts $1$ and $2$ denote the nearest and second-nearest clusters for each coefficient, respectively. The fuzzy clustering FIR architecture implement is presented in Fig.\ref{fig:fuzzy cluster FIR architecture.}. 

\begin{figure*}[htbp]
  \centering
  \includegraphics[width=\linewidth]{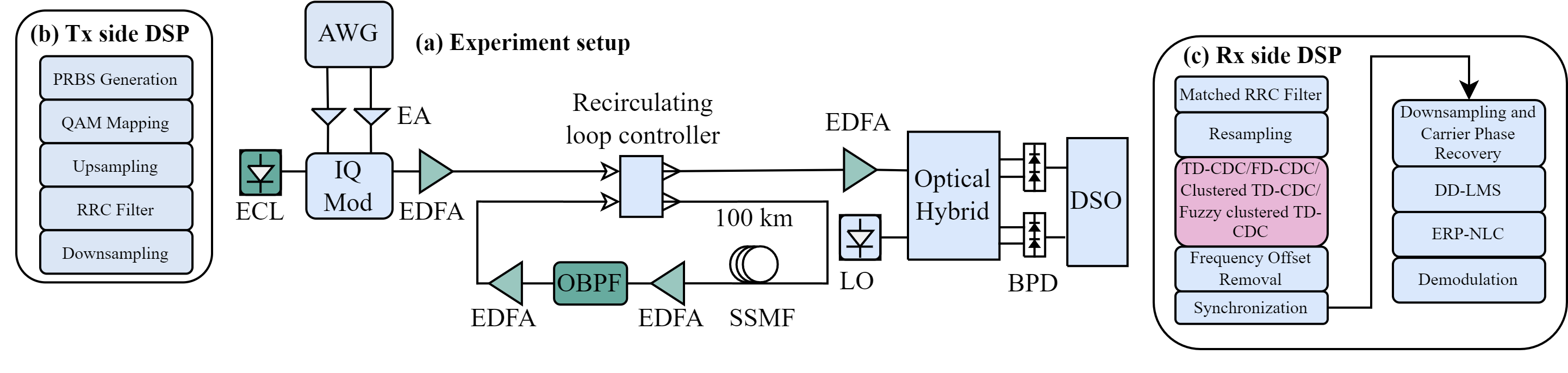}
  \caption{(a) The experiment setup of the 20baud single-channel single-polarization 16-QAM 18 $\times$ 100 km coherent optical communication system. ECL: external cavity laser, AWG: arbitrary waveform generator, EA: electrical amplifiers, EDFA: erbium-doped fiber amplifier, OBPF: optical band-pass filter, LO: local oscillator, BPD: balanced photodetector, DSO: digital sampling oscilloscope. (b) Tx-side DSP; (c) Rx-side DSP. TD: time domain; FD: frequency domain; DD-LMS: decision-directed least mean square filter; ERP-NLC: enhanced regular perturbation-based nonlinearity compensation.}
  \label{fig:Exp setup and DSP stack.}
\end{figure*}

\paragraph*{\large Proof-of-concept experiment.} The experimental setup for 20Gbaud single-channel single-polarization (SP) 16-QAM coherent transimission over 18 $\times$ 100 km of standard single-mode fiber (SSMF) is illustrated in Fig.\ref{fig:Exp setup and DSP stack.}(a). At the transimitter, an external cavity laser (ECL, ID Photonics CBMX-4-CH01, with a linewidth of 100 kHz) operating at a central wavelength of 1550 nm serve as the optical source. The 16-QAM signal produced by an arbitrary waveform generator (AWG, Tektronix AWG70002, with a 3-dB bandwidth of 10 GHz) is amplified using a pair of electrical amplifiers (EA, SHF M834 B) to drive the IQ modulators(Fujitsu, FTM7962EP). An erbium-doped fiber amplifier (EDFA, MCEDFA-LA-17-1-FA-T1) is employed to regulate the launch power prior to the fiber loop. A recirculating loop controller (BRIMROSE, AMM-100-4-140-C-RLS(nfs)-RM) utilizing acousto-optics modulators is implemented to simulate the long-haul fiber transimission link, incorporating a 100 km ring of SSMF and EDFAs. An optical band-pass filter (OBPF, Alnairlabs BVF-200CL) is applied to mitigate the accumulation of amplified spontaneous emission (ASE) noise within the loop. At the receiver, the signal is initially amplified by an EDFA to adjust the received power. Subsequently, the signal and the local oscillator (LO) are directed into the optical hybird. The down-converted electrical signal is amplified and recorded by a digital sampling oscilloscope (DSO, Tektronix DPO 72504DX) with a sampling rate of 50GSa/s to perform offline digital signal processing (DSP).

The DSP diagram is shown in Fig.\ref{fig:Exp setup and DSP stack.}(b) and \ref{fig:Exp setup and DSP stack.}(c). In the Tx-side DSP, the generated pseudo-random binary sequence (PRBS) bitstream is mapped to 16-QAM symbols, followed by digital shaping using a root-raised cosine (RRC) filter with a roll-off factor of 0.1. In the Rx-side DSP, a matched RRC filter is employed to mitigate inter-symbol interference (ISI). The 16-QAM symbols are then re-sampled to two samples per symbol. Following this, chromatic dispersion compensation is applied. After the frequency offset removal, synchronization, down-sampling, carrier phase recovery, and a decision-directed least mean square (DD-LMS) filter is utilized to equalize the remained linear impairments. The enhanced regular perturbation-based nonlinearity compensation (ERP-NLC) technique is subsequently implemented to counteract the fiber Kerr nonlinear effects \cite{kumar2019enhanced}. Finally, the 16-QAM symbols are de-modulated, BER and Q-factor are computed, with the Q-factor derived from the BER using the formula $Q = \sqrt{2}\text{erfc}^{-1}(2\text{BER})$.

\paragraph*{\large Results and discussions.}
To evaluate the effectiveness of our proposed method, we employed a state-of-the-art frequency-domain equalizer based on FFT as a benchmark for chromatic dispersion compensation. The complexity estimation for the TD-CDC\cite{savory2008digital,xu2010chromatic}, clustered TD-CDC \cite{gomes2024geometric}, fuzzy clustered TD-CDC, and frequency-domain CDC (FD-CDC) is quantified by the number of real multiplcations per equalized symbol (RMPS). All complex-valued multiplcations are implemented using Karastuba algorithm \cite{weimerskirch2006generalizations}, in which each complex multiplcation is decomposed into 3 real-valued multiplications. Therefore, the complexity of TD-CDC by using FIR filtering is given by $C = 3(N - 1) / 2$, where $N$ is the filter length. However, for the clustered TD-CDC, which utilize a cluster filter, the complexity is expressed as $C = 3 N_c$, where $N_c$ denotes the number of clusters employed in the filter. Since the weights $\alpha$ and $1 - \alpha$ can be pre-stored in a look-up table, the complexity of proposed fuzzy clustered TD-CDC is equivalent to that of clustered TD-CDC (i.e. $C = 3 N_c$). In the case of FD-CDC, assuming a radix-2 algorithm, the number of RMPS is estimated as:
\begin{equation}
  C = N_{\text{FFT}}\frac{3 \log_2(N_{\text{FFT}}) + 3}{N_{\text{FFT}} - N_{\text{Overlap}} + 1}
\end{equation}
in which $N_{\text{FFT}}$ represents the FFT size, $N_{\text{Overlap}}$ denotes the required overlap size which is designated as half of the FFT size.

\begin{figure*}
  \centering
  \includegraphics[width=0.75\linewidth]{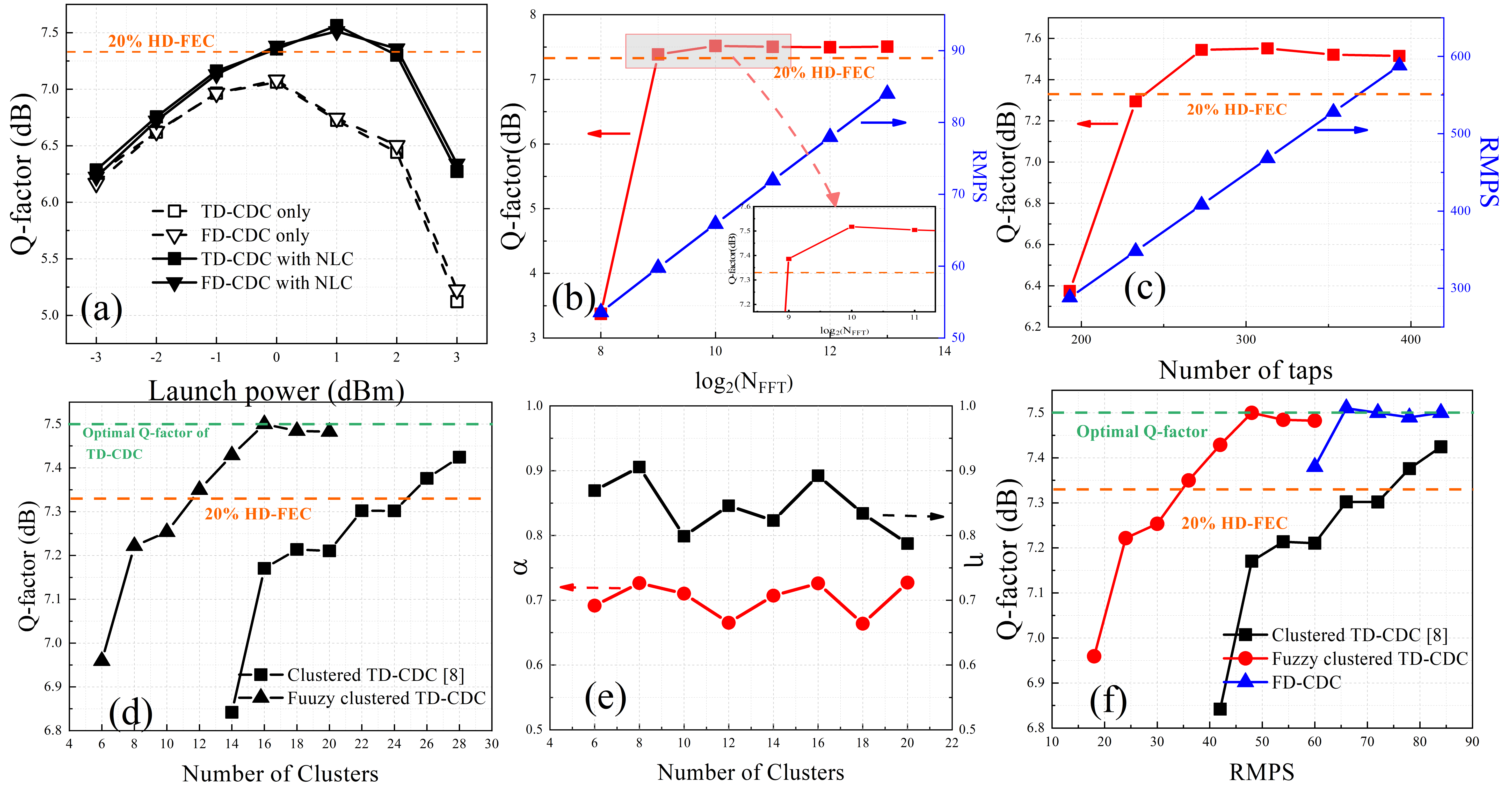}
  \caption{(a) The Q-factor versus launch powers for time-domain and frequency-domain chromatic dispersion compensation schemes under unconstrained complexity; NLC: nonlinearity compensation. (b) The Q-factor and complexity versus different FFT sizes for frequency-domain chromatic dispersion compensation scheme; RMPS: real multiplcation per recoverd symbol. (c). The Q-factor and complexity versus different number of taps for time-domain chromatic dispersion compensation scheme. (d) Performance and complexity for different cluster quantities for clustered and fuzzy-clustered time-domain chromatic dispersion compensation schemes. (e) The optimized hyperparameters $\alpha$ and $\eta$ for fuzzy clustered TD-CDC versus the number of clusters. (f) The Q-factor as a function with complexity for different chromatic dispersion compensation schemes.}
  \label{fig:results} 
\end{figure*}

%Fig.1
First, we analyze the performance of TD-CDC and FD-CDC at different launch powers under unconstrained complexity. For the TD-CDC, the filter length $N$ is set to 393, while for FD-CDC, the FFT size is 2048. As shown in Fig.\ref{fig:results}(a) that both schemes achieve identical Q-factor. Furthermore, with ERP-NLC, the optimal launch power is 1.0 dB higher than that of CDC-only, and the best Q-factor improves by $\sim$ 0.5 dB. A 20\% hard-decision forward error correction (HD-FEC) threshold of 1e-2 is adopted as the target BER threshold, corresponding to a Q-factor of 7.33 dB. In the following section, we assess the performance and complexity of various chromatic dispersion compensation schemes at the optimal launch power of 1.0 dBm.

% Fig.2
Fig.\ref{fig:results}(b) illustrates the performance and complexity of FD-CDC versus FFT size. The Q-factor improves with increasing FFT size before saturating at an FFT size of 1024. At the target Q-factor of 20\% HD-FEC, the minimum required FFT size for FD-CDC is 512, with a complexity of 60 RMPS. 
Similarly, the performance and complexity of TD-CDC versus filter taps is investigated, as shown in Fig.\ref{fig:results}(c).The Q-factor for TD-CDC improves with increasing taps before approaching its optimal value. For TD-CDC, the minimum required number of taps at the target Q-factor of 20\% HD-FEC is approximately 273, with a complexity of 408 RMPS. The complexity of traditional TD-CDC is more than 6 times that of FD-CDC, highlighting the need for a low-complexity time-domain chromatic dispersion compensation scheme. Clustered TD-CDC in Ref. \cite{gomes2024geometric} employs hard decision to group the tap coefficients to the nearest centroid, thereby reducing calculation complexity. However, the error between the initial coefficient and its associated centroid inevitably degrades system performance, limiting the transimission fiber length to 480 km.

Fuzzy clustering can correct the error caused by hard decision clustering. Fig.\ref{fig:results}(d) shows the Q-factor versus the number of clusters for fuzzy clustered TD-CDC and clustered TD-CDC when the number of filter taps is 273. 
At the target Q-factor of 20\% HD-FEC, the minimum required number of clusters, $N_c$, is 26 for clustered TD-CDC and 12 for fuzzy clustered TD-CDC, resulting in a reduction of over 50\% in the number of clusters, which is proportional to the RMPS. Furthermore, compared to clustered TD-CDC, the fuzzy clustered TD-CDC scheme exhibits a faster improvement in Q-factor as the number of clusters increases, and it can approach the optimal Q-factor of unclustered TD-CDC with a relatively small number of clusters (i.e. $N_c = 16$).
The hyperparameters $\eta$ and $\alpha$ of fuzzy clustered TD-CDC are optimized using Bayesian Optimization to maximize the Q-factor. The corresponding hyperparameters for each number of clusters are show in Fig.\ref{fig:results}(e), where both the optimized $\alpha$ and $\eta$ fluctuate only slightly for each different clusters. For an intuitive comparsion with FD-CDC, we present the Q-factor and complexity of different schemes in Fig.\ref{fig:results}(f). If 20\% HD-FEC is set as a target, the RMPS is 36, 78, and 60 for fuzzy clustered TD-CDC, clustered TD-CDC, and FD-CDC, respectively. Fuzzy clustered TD-CDC has complexity savings of 53.8\% and 40\% with comparsion to clustered TD-CDC and FD-CDC.
Moreover, fuzzy clustered TD-CDC achieves the same optimal Q-factor as FD-CDC while lowering the minimum required complexity by 27\%. These results experimentally demonstrate the effectiveness of our proposed fuzzy clustered TD-CDC in achieving a superior performance-complexity trade-off.

\paragraph*{\large Conclusion} In this letter, we present a novel two-stage fuzzy clustered time-domain chromatic dispersion compensation scheme for long-haul coherent optical communication systems. The effectiveness of the proposed method is validated through experiments on a single-channel, single-polarization 16-QAM, 20Gbaud, 18$\times$ 100 km coherent optical fiber transimission system. First, the centroids of filter coefficients are determined using the K-means clustering algorithm. Then, a fuzzy soft decision is applied to the coefficient points based on the Euclidean distance between the points and the centroids in the complex plane. Experimental results demonstrate that, compared to FD-CDC and clustered TD-CDC, the proposed method reduces the RMPS at a target Q-factor of 20\% HD-FEC by 40\% and 53.8\%, respectively. Moreover, the proposed mehtod can achieve optimal performance as FD-CDC with a 27\% reduction in RMPS. 

\begin{backmatter}
\bmsection{Funding} National Natural Science Foundation of China (61427813).

\bmsection{Disclosures} The authors declare no conflicts of interest.

\bmsection{Data Availability Statement} Data underlying the results presented in this paper are not publicly available at this time but may be obtained from the authors upon reasonable request.
\end{backmatter}

% Bibliography
\bibliography{sample}
\bibliographyfullrefs{sample}

\end{document}